\pdfoutput=1
\documentclass[11pt]{article}

\usepackage[final]{acl}  

\usepackage{times}
\usepackage{latexsym}
\usepackage[T1]{fontenc}
\usepackage[utf8]{inputenc}
\usepackage{microtype}
\usepackage{inconsolata}
\usepackage{tabularx}
\usepackage{booktabs}

\usepackage{graphicx}
\usepackage{soul}
\usepackage{subcaption}

\title{Emoji Reactions on Telegram:\\Unreliable Indicators of Emotional Resonance} 

\author{
  \textbf{Serena Tardelli\textsuperscript{1}},
  \textbf{Lorenzo Alvisi\textsuperscript{1,2}},
  \textbf{Lorenzo Cima\textsuperscript{1,3}},\\
  \textbf{Stefano Cresci\textsuperscript{1}},
  \textbf{Maurizio Tesconi\textsuperscript{1}} \\
  \textsuperscript{1}Institute of Informatics and Telematics, National Research Council, Italy \\
  \textsuperscript{2}IMT School for Advanced Studies Lucca, Italy \\
  \textsuperscript{3}Department of Information Engineering, University of Pisa, Italy \\
  \texttt{\{firstname.lastname\}@iit.cnr.it}
}

\begin{document}
\maketitle
\begin{abstract}
Emoji reactions are a frequently used feature of messaging platforms, yet their communicative role remains understudied. Prior work on emojis has focused predominantly on in-text usage, showing that emojis embedded in messages tend to amplify and mirror the author's affective tone. This evidence has often been extended to emoji reactions, treating them as indicators of emotional resonance or user sentiment. However, they may reflect broader social dynamics. Here, we investigate the communicative function of emoji reactions on Telegram. We analyze over 650k crypto-related messages that received at least one reaction, annotating each with sentiment, emotion, persuasion strategy, and speech act labels, and inferring the sentiment and emotion of emoji reactions using both lexicons and LLMs. We uncover a systematic mismatch between message and reaction sentiment, with positive reactions dominating even for neutral or negative content. This pattern persists across rhetorical strategies and emotional tones, indicating that emojis used as reactions do not reliably function as indicators of emotional mirroring or resonance of the content, in contrast to findings reported for in-text emojis. Finally, we identify the features that most predict emoji engagement. Overall, our findings caution against treating emoji reactions as sentiment labels, highlighting the need for more nuanced approaches in sentiment and engagement analysis.
\end{abstract}

\section{Introduction}
Emoji reactions are a central feature of digital platforms, enabling users to respond quickly and visually to content. While much research has focused on in-text emojis (i.e., linguistic elements embedded within authored messages), 
emoji reactions (i.e., reactions appended by readers as lightweight feedback mechanisms), remain comparatively underexplored. Yet they are equally revealing, as they are instantaneous, costless, and a quick way to interact with content without composing a reply.
Moreover, while prior work has examined in-text emojis, showing they amplify sentiment~\cite{shiha2017effects}, improve sentiment classification~\cite{liu2021improving}, and provide emotional nuance~\cite{lou2024emoji, khan2025sentiment}, this understanding has often been extended implicitly to emoji reactions~\cite{pool2016distant}, treating them as affective signals. However, recent work suggests that emoji reactions may instead express pragmatic or social meanings, such as to approve or normalize toxic discourse~\cite{morales2025thumbs}. %

Building on this idea, we provide the first large-scale study examining whether emoji reactions actually reflect emotional resonance, an assumption often implicitly adopted when reactions are used as proxy labels in sentiment analysis. To do so, we investigate the communicative function of emoji reactions on Telegram, a messaging platform widely used for news, politics, and activism. On Telegram, emoji reactions are publicly visible and frequently used, making them a rich signal of collective response. In detail, we analyze over 650k crypto-related Telegram messages and their associated emoji reactions. We combine large-scale emotion classification with rhetorical analysis to examine the alignment between message tone and reaction type.
In particular, we address three research questions:
\begin{itemize}
    \item \textbf{RQ1}: Do emoji reactions align with the sentiment expressed in the original message?
    \item \textbf{RQ2}: Do emoji reactions reflect the emotional tone of the original message, creating emotional resonance? 
    \item \textbf{RQ3}: Which message features predict emoji reactions? 
\end{itemize}

We adopt a mixed-methods approach to examine messages and their reactions across multiple dimensions, including sentiment, emotional tone, persuasion strategy, and speech act. 
We find that positive emoji reactions dominate, even in response to messages that are emotionally neutral or negative. This consistent mismatch holds across rhetorical strategies and emotional categories, suggesting a pragmatic function for some emoji reactions, more closely tied to social approval or support than to shared emotional resonance. 
Finally, we show that certain rhetorical strategies are strong predictors of emoji engagement, and that emotionally neutral or negative messages tend to attract more emoji reactions overall. Our contributions are manifold:
\begin{itemize}
    \item We provide the first large-scale analysis of emoji reactions on Telegram to determine whether they distort or amplify emotional alignment.
    \item We conduct this analysis on a large dataset of 650k crypto-related Telegram channels, a domain known for strong community identity and high engagement, where reactions are especially frequent and publicly visible.
    \item We show a consistent misalignment between message sentiment and emoji reactions, indicating that, unlike in-text emojis, reactions do not reliably act as emotional mirroring.
    \item We identify specific message features, such as emotional framing or persuasive tactics, that predict user engagement in terms of emoji reactions.
\end{itemize}

Our findings suggest that emoji reactions are not reliable proxies for sentiment content, as they weakly mirror or resonate with the emotion evoked by the message. 
Together, these results challenge the assumption that emoji reactions reliably reflect emotional response, as commonly observed for in-text emojis. 
These patterns should also be interpreted in light of Telegram's specific communicative context, where community norms may shape reaction behavior.
More broadly, these findings have important implications, particularly for misinformation and political communication research, where reaction-based signals are often interpreted as indicators of emotional resonance or public sentiment.

%

\section{Related work}

Emojis have been extensively studied as markers of affective communication in digital discourse. Early work recognized their role as nonverbal cues that supplement text with emotional tone~\cite{miller2016blissfully,kralj2015sentiment}, and many researchers have treated them as reliable proxies for emotion. For instance, work in~\cite{shiha2017effects} found that emoji amplify sentiment in tweets. Work in~\cite{liu2021improving} showed that treating emojis as features can significantly improve sentiment classification accuracy. Work in~\cite{lou2024emoji} showed that emojis provide emotional context in posts highlighting the role of emojis as emotional indicators. Similarly, work in~\cite{khan2025sentiment} showed that emojis play a vital role in sentiment expression, often conveying more explicit and nuanced information than the text alone.
As a result, datasets such~\cite{kralj2015sentiment}, which assigns sentiment scores to individual emojis based on crowdsourced annotations, has been widely used for downstream sentiment classification. Similarly, the DeepMoji model~\cite{felbo2017using} was trained on tweets labeled with emojis to learn emotional representations, relying on the assumption that emoji usage reflects the emotional intent of the author.

This assumption has been extended to emoji reactions, predefined emoji responses users can select to express a reaction to content, on platforms like Facebook. These reactions are often interpreted as ground truth labels for emotion classification tasks~\cite{pool2016distant}, enabling models to learn emotion categories from reaction distributions. However, more recent studies caution against this approach. Work in~\cite{graziani2019jointly} observed that reactions can be semantically ambiguous or perform multiple pragmatic functions. Authors in~\cite{paolillo2023awkward} notes that reactions can operate not only as affective signals but as interactional cues, marking approval, affiliation, or irony depending on context. 

Our work extends these critiques by examining emoji reactions on Telegram, a platform where reactions are public, optional, and highly community-driven. Unlike prior research that treats emoji reactions as emotional response, we find systematic mismatch between the sentiment expressed in a message and the sentiment inferred from emoji reactions. 
Messages with negative or neutral tone frequently receive positive emoji reactions, such as the thumb up or love emoji. This suggests that reactions function more as social endorsement than as expressions of shared affect. 
The result challenges the methodological practice of using reaction emojis as emotional supervision signals, at least in environments where group identity and approval play a stronger role than emotion mirroring. 

On a similar line, recent work~\cite{morales2025thumbs} explored how emoji reactions are used in toxic versus non-toxic contexts on Telegram, showing that even toxic comments often receive positive emoji reactions, signaling approval or normalization of harmful content~\cite{cima2025investigating}. Our findings complement and generalize this pattern: while~\citet{morales2025thumbs} focused specifically on toxicity, we observe that positive reaction bias holds across a broader range of communicative intentions, including persuasive tactics, emotional tone, and speech acts. Moreover, while our results do not exclude the possibility of emotional mirroring, they suggest that social signaling frequently overrides it in practice.

Further, prior work on emotional dynamics in engagement showed that emotionally charged content tends to propagate faster~\cite{vosoughi2018spread}, and that moral-emotional language increases virality within ideologically homogeneous groups~\cite{brady2017emotion}. Our findings complement this literature by showing that the type of emotional response conveyed through reactions is not always aligned with message tone, and instead may reinforce in-group signaling mechanisms. This distinction between emotion expression and interactional feedback is especially relevant for studying engagement in polarized or activist spaces, where affective reactions might serve to signal allegiance more than to reflect emotional resonance.

By focusing on Telegram, which remains comparatively understudied despite its growing role in political discourse and information diffusion, we contribute new evidence on communication dynamics within this platform. While recent studies have examined toxicity, misinformation, and content moderation on Telegram~\cite{urman2022they,alvisi2024unraveling,alvisi2025mapping}, the communicative function of emoji reactions remains largely unexplored, a gap that we address in the present study.
\section{Dataset}

We collect our dataset by retrieving Telegram channels from TGStat\footnote{\url{https://tgstat.com/}}, a third-party analytics platform that categorizes public channels by topic, language, and popularity. We select only English-language channels with at least 5,000 subscribers. This initial sampling yields 20,695 channels. For each channel, we collect all public messages posted between January and December 2024, and focus on those that received at least one emoji reaction. 
The resulting dataset comprises 647,879 messages from 993 chats related to the topic of cryptocurrencies. Message distribution follows a long-tail pattern consistent with a power law, with few channels accounting for most messages. While Telegram supports both default and custom emojis for reactions, our analysis includes only standard emojis.

\section{Methods}

We analyze Telegram messages using a multi-stage pipeline that combines NLP-based linguistic annotation with behavioral cues derived from user reaction data (i.e., emoji-based reactions). First, we automatically annotate each message with linguistic features, including  sentiment, persuasion strategies, emotion, and speech act categories. Then, we annotate the emoji reactions to infer both their sentiment polarity and emotional content.
This allows capturing both linguistic content and social responses. 

\paragraph{Message Annotation}
We enrich each Telegram message with linguistic and psychological features through multiple classification layers.
\begin{itemize}

    \item \textit{Persuasion strategies}: We detect rhetorical appeals of each message (e.g., emotion, authority, scarcity, social proof, fear, popularity, logic, reciprocity, fear of missing out) plus a neutral and a no persuasion categories using the \textit{valhalla/distilbart-mnli-12-1} zero-shot model~\cite{valhalla2020distilbartmnli} and the taxonomy by~\citet{qachfar2023redaspersuasion}.

    \item \textit{Emotion analysis}: We assign each message an emotion from~\cite{plutchik1994psychology}'s taxonomy using the \textit{bhadresh-savani/bert-base-uncased-emotion} model~\cite{bhadresh2021emotionbert}. 

    \item \textit{Sentiment analysis}: We label messages as positive, neutral, or negative using \textit{cardiffnlp/twitter-roberta-base-sentiment}~\cite{cardiffnlp2021sentimentroberta}, widely adopted in computational social science~\cite{jahin2024hybrid}.
    
    \item \textit{Speech acts}: We classify messages as assertives, directives, commissives, expressives, and declarations based on~~\citet{searle1969speech} taxonomy~\cite{jegede2024speech,saha2020transformer} using the \textit{valhalla/distilbart-mnli-12-1} zero-shot model~\cite{valhalla2020distilbartmnli}.

\end{itemize}

\paragraph{Emoji Reaction Interpretation}
We infer the emotional and affective content of reactions using two complementary methods:
\begin{itemize}
    \item \textit{Emoji sentiment score}: We use the emoji sentiment lexicon~\cite{kralj2015sentiment}\footnote{\url{https://www.clarin.si/repository/xmlui/handle/11356/1048}}, which assigns each emoji a polarity ($-1$, $0$, $+1$). For each message, we compute a weighted average score ($-1$ to $+1$) based on reaction frequencies, then map it to positive ($\ge0.2$), negative ($\le-0.2$), or neutral, as done in prior work~\cite{chifu2015web,chaithra2019hybrid} showing that defining a neutral range around zero reduces label noise and improves the robustness of downstream sentiment analyses. This mapping allows us to assign a discrete sentiment label to the emoji reaction set for each message.
    \item \textit{LLM-based emoji emotion}: We prompt OpenAI GPT-4o model with the list of reaction emojis to assign an emotion from~\citet{plutchik1994psychology}'s taxonomy (coarse analysis) or a fine-grained emotional label, allowing for more nuanced and free interpretation beyond the core categories. In this way, we can also assign each message the emotion of its most frequent reaction emoji.
\end{itemize}

\paragraph{Predictive Modeling and Feature Interpretation}
We investigate which aspects of a message predict its likelihood of receiving a large number of emoji reactions. We frame the task as a binary classification problem. We train a binary Random Forest classifier with 28 features to distinguish between ``high engagement'' posts and ``low engagement'' ones, using the median number of emoji reactions as threshold. While the classification task is not the core goal, this predictive setup allows us to quantify the impact of individual message-level features. To understand the contribution of each feature to engagement, we leverage SHAP, a game-theoretic method that has become standard in explainable AI~\cite{lundberg2017unified}. 
SHAP assigns each feature a local importance score for every prediction, allowing us to rank linguistic and emotional cues according to their predictive power, as done in multiple related social media analysis tasks~\cite{gambini2024anatomy, tessa2024beyond}.

\section{Results}

\subsection{RQ1 - Alignment of emoji reactions with sentiment}

\begin{figure}[t]
    \centering
    \includegraphics[width=0.75\columnwidth]{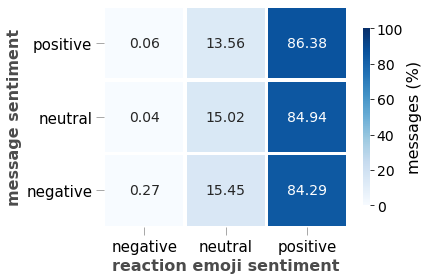}
    \caption{Message sentiment \textit{vs} emoji reactions sentiment. Positive emoji reactions dominate independently of message sentiment.}
    \label{fig:sent_vs_sent}
\end{figure}

\begin{figure*}[th!]
  \centering
  \begin{subfigure}[t]{0.48\linewidth}
    \centering
    \includegraphics[width=0.97\linewidth]{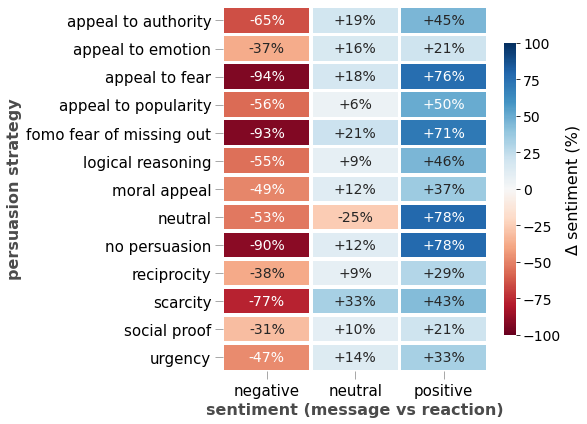}
    \caption{Message sentiment \textit{vs} emoji reactions sentiment, for different persuasive strategies. Most strategies show a consistent shift toward more positive sentiment in reactions. Sentiment distributions in messages and reactions differ significantly across all persuasion strategies (Chi-squared test, $p<0.001$), revealing a systematic misalignment.}
    \label{fig:delta_pers}
  \end{subfigure}%
  \hfill
  \begin{subfigure}[t]{0.48\linewidth}
    \centering
    \includegraphics[width=0.71\linewidth]{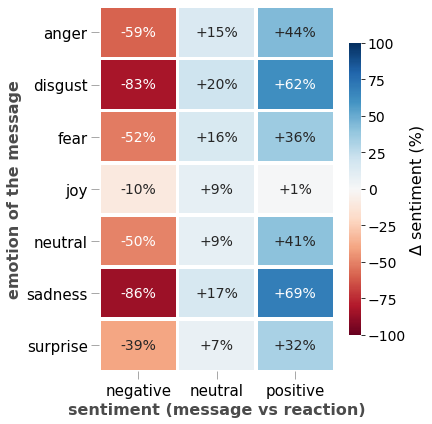}
    \caption{Message sentiment \textit{vs} emoji reactions sentiment, for different emotions. Negative emotions (e.g., sadness, disgust) receive positive reactions. Sentiment distributions in messages and reactions differ significantly across all emotions (Chi-squared test, $p<0.001$).}
    \label{fig:delta_emot}
  \end{subfigure}
  \caption{Sentiment shift from messages to reactions by persuasion strategy (left) and emotion category (right).}
  \label{fig:fig-rq1s}
\end{figure*}

In Figure~\ref{fig:sent_vs_sent} we compare the sentiment conveyed in the message with the sentiment of the emoji reactions it receives. The majority of messages, regardless of whether they express positive, neutral, or negative sentiment, receive predominantly positive reactions. For instance, over 84\% of negative and neutral messages are met with positive emoji sentiment, with less than 0.3\% of negative messages receiving negative emoji feedback. This suggests a systematic mismatch between the emotional tone of the message and the sentiment expressed in the reactions, an inconsistency in line with previous work on other platforms~\cite{wang2023more}.

To further explore this phenomenon, in Figure~\ref{fig:fig-rq1s} we break down the analysis by persuasion strategy and emotion detected in the message. In both cases, we plot the delta in sentiment (i.e., the difference between the sentiment of the message and the weighted sentiment of its emoji reactions). These differential heatmaps reinforce the pattern we previously observed. While the original sentiment expressed in messages spans a wide range, the received sentiment, as inferred from emojis, is skewed toward positive or neutral categories. This suggests a general positivity bias in emoji usage, potentially reflecting users' preference for agreement, support, or acknowledgment over dissent.

In Figure~\ref{fig:delta_pers}, this effect holds across most persuasive strategies. Techniques like appeal to fear, fear of missing out, and scarcity, which typically rely on negatively framed content, still elicit predominantly positive emoji reactions, with net sentiment shifts exceeding $+70$\%. However, the magnitude of this shift varies across strategies. For instance, techniques such as reciprocity, fear of missing out, or appeal to emotion show more balanced distributions, with a smaller difference between message tone and reaction sentiment. This suggests that not all forms of persuasion equally trigger endorsement, and that emotional proximity between message and reaction may play a role in moderating the shift. A similar pattern emerges in Figure~\ref{fig:delta_emot}, where messages expressing sadness, disgust, or fear are still met with disproportionately positive reactions. However, the shift is not uniform, as messages labeled with joy or surprise show only modest changes, indicating that positive messages are not necessarily amplified through emoji reactions. Interestingly, this tendency is especially evident in response to emotionally neutral content, where reactions may reflect a desire to affirm or engage with others, convey presence, support, or as a form of interpersonal acknowledgment.

\subsection{RQ2 - Emoji reactions and emotional resonance}
\begin{figure}[t]
\centering
  \includegraphics[width=\columnwidth]{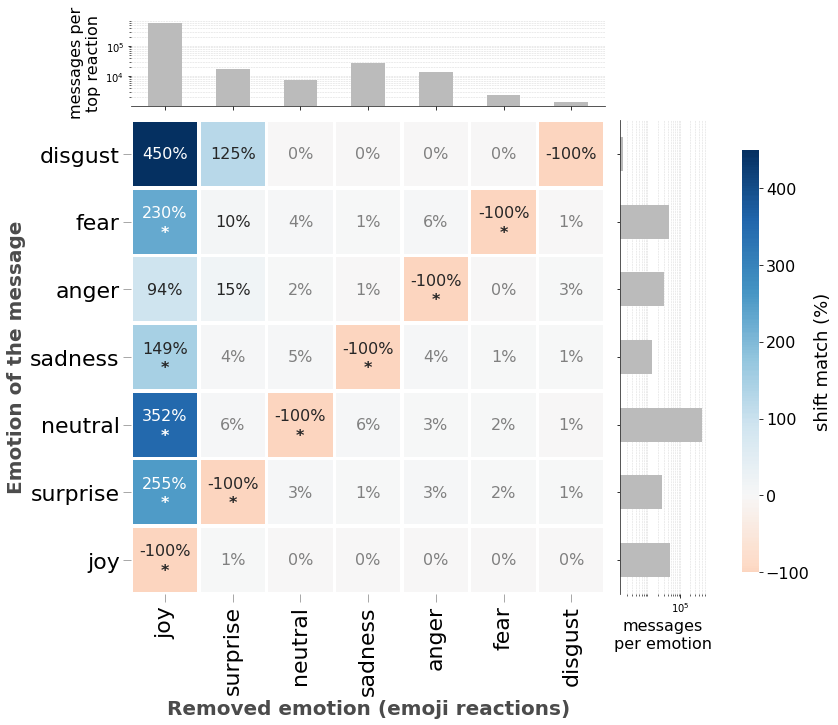}
  \caption{Percentage change in alignment between message emotions and top emoji reactions after removing each emotion. Removing joy leads to large and statistically significant improvements.}
  \label{fig:toprank_improv}
\end{figure}

While RQ1 revealed a tendency for messages to receive positive emoji reactions regardless of their sentiment, the analysis was based on a coarse-grained polarity framework (positive, neutral, negative). To better understand which emotions drive this misalignment, RQ2 shifts focus from sentiment to discrete emotion categories. We ask whether certain emoji reactions systematically distort alignment with the emotional tone of the message, and whether removing them improves interpretability. 
Specifically, we hypothesize that emoji reactions are not always used to mirror the emotional tone of a message, but instead often serve as signals of social approval, engagement, or banter. To evaluate this, we measure how often the emotion of the message appears among the top-ranked emoji reactions.
In other words, we compute the percentage of messages where the top-ranked emoji reaction shares the same emotion category as the message itself.
To assess the impact of individual emoji reactions, we perform an ablation analysis. We remove one emotion at a time from the reaction distribution and we re-calculate the top-ranked emoji reaction. If removing a specific emotion significantly increases the number of matches (i.e., if it makes the dominant emoji reaction more likely to align with the emotion of the message), we interpret that category as disruptive to emotional alignment (i.e., the removed reaction was often selected even when it did not match the emotional tone of the message). Figure~\ref{fig:toprank_improv} presents the relative change in alignment match after removing each emoji reaction category, grouped by the emotion label of the message. Each cell shows the percentage change in the proportion of messages where the emotion of the top-ranked emoji matches the emotion of the message. Statistically significant changes, defined as those with a standardized effect size (Cohen's h) $\ge 0.2$, are marked with an asterisk. Cohen's h quantifies the magnitude of difference between proportions before and after ablation, providing an interpretable measure of effect strength beyond percentages.

As expected, the lowest scores appear along the diagonal, where the removed emoji category matches the labeled emotion of the message. In fact, removing the correct category naturally lowers alignment, simply because it eliminates true matches. In these cases, the drop does not indicate interference from other emotions, but simply reflects the expected loss in alignment from removing the correct category. The main takeaway from our results is that removing joy consistently leads to large and statistically significant improvements in alignment, especially for messages labeled as sadness, neutral, surprise, and fear. In contrast, removing other reaction emotions has minimal impact. For instance, while removing surprise reactions results in a large relative increase in matches for disgust messages (i.e., +125\%), the number of messages labeled with disgust is very small, making this effect statistically negligible.

The consistent improvement in alignment when joy reactions are removed suggests that these emojis are often used even when they do not match the emotional tone of the message. Unlike other categories, joy appears to systematically distort emotional alignment, pointing to a distinct functional role. Rather than reflecting shared affect, joy reactions may serve pragmatic or social purposes, such as signaling encouragement, approval, banter, or support, regardless of the emotional content of the message. 
The disruptive effect of joy, revealed through targeted ablation, highlights its unique communicative function in digital contexts, one that is orthogonal to emotional resonance.

Our previous analysis relied a binary metric: whether the top-ranked emoji reaction matched the emotion expressed in the message. While this provides an intuitive signal of alignment, it reduces the full distribution of emoji reactions to a binary outcome. 
To capture a more nuanced and robust measure of alignment, we compute the cosine similarity between the distribution of emoji reaction emotions and the message emotion represented as a one-hot vector. This allows us to detect subtle changes in alignment, not just categorical shifts. Unlike the previous binary approach, cosine similarity reflects the degree to which the overall emotional tone of emoji reactions matches the intended emotion of the message, making it more sensitive and interpretable. This is particularly important when testing whether the presence of specific emoji emotions obscures rather than reflects affective resonance.
To identify which emotions most affect this alignment, we systematically removed each emotion category from the emoji distribution and measured the resulting change in cosine similarity ($\Delta$). A negative effect indicates that removing the emotion improves alignment, suggesting that its presence may distort the emotional signal conveyed by emoji reactions.

\begin{figure}[t]
    \centering
    \includegraphics[width=0.75\columnwidth]{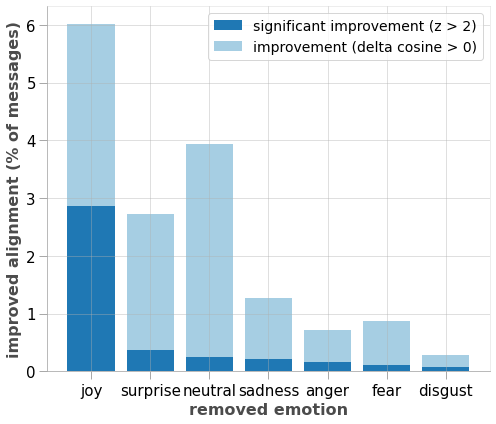}
    \caption{Percentage of messages with improved alignment after removing each emoji reaction emotion. Joy most frequently obscures alignment between message and reaction tone.}
    \label{fig:cosine_improv}
\end{figure}

Among all categories, joy produced the most pronounced negative effect ($\Delta=-0.107$), followed by surprise, with a much smaller change ($\Delta=-0.006$). Other emotions, such as disgust ($\Delta=-0.0007$), had negligible impact. This suggests that joy-related emojis, more than any other category, systematically misalign with the emotional tone of the message. We then examined, for each emotion removed from the emoji reaction distribution, how often this ablation led to improved alignment with the emotion of the message. Figure~\ref{fig:cosine_improv} shows the results. While all emotions showed low rates of improvement in cosine similarity (i.e., $<4$\%, with most $<1$\%), removing joy led to a substantial increase in alignment in over $6$\% of the messages. Moreover, joy accounted for over $17,000$ statistically significant improvements (z-score $>2$), surpassing any other emotion. This pattern is not explained simply by frequency. While joy appears in over 95\% of emoji reactions, its removal leads to disproportionately many and significantly large alignment improvements, suggesting that joy plays a qualitatively different role. In fact, while sadness appears in over 26\% of emoji reactions, removing it leads to improved alignment in just 1.27\% of messages, compared to 6.02\% for joy. No other emotion, regardless of frequency, produces such a consistently large and disproportionate improvement in alignment when removed. These findings reinforce the idea that joy reactions are not merely emotional mirroring, but often mask or override the underlying sentiment of a message. 


\begin{figure}[t]
    \centering
    \includegraphics[width=0.8\columnwidth]{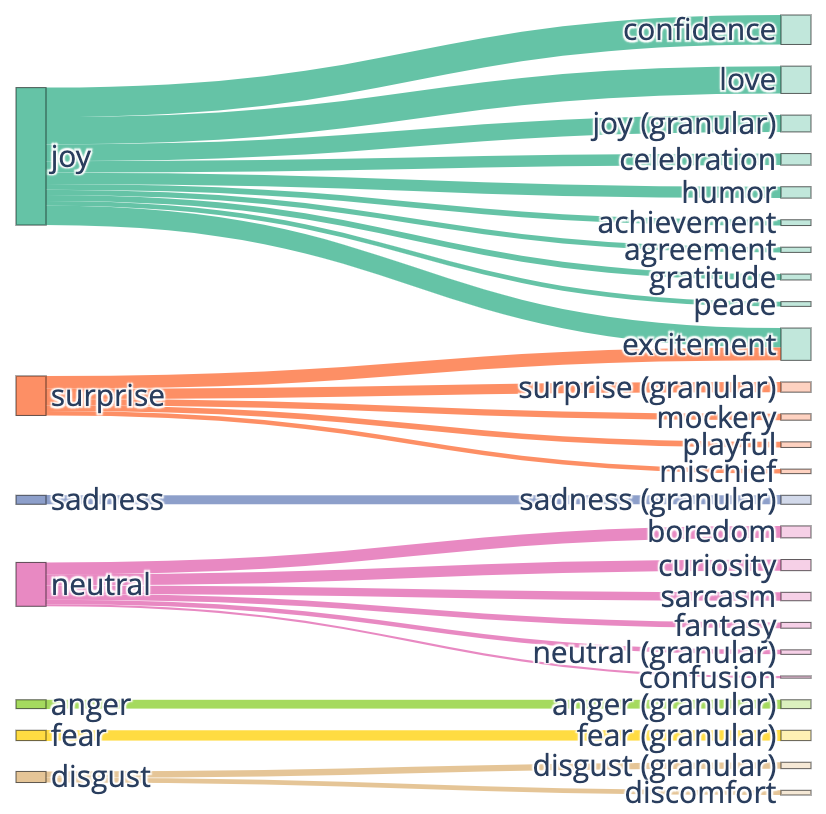}
    \caption{Mapping of coarse-grained emotions (left) to their finer-grained subcategories (right). Even when emojis reflect joy or neutrality, they frequently encode more specific sentiments of approval like agreement, gratitude, and excitement.}
    \label{fig:sankey}
\end{figure}

\begin{figure}[t]
    \centering
    \includegraphics[width=0.85\columnwidth]{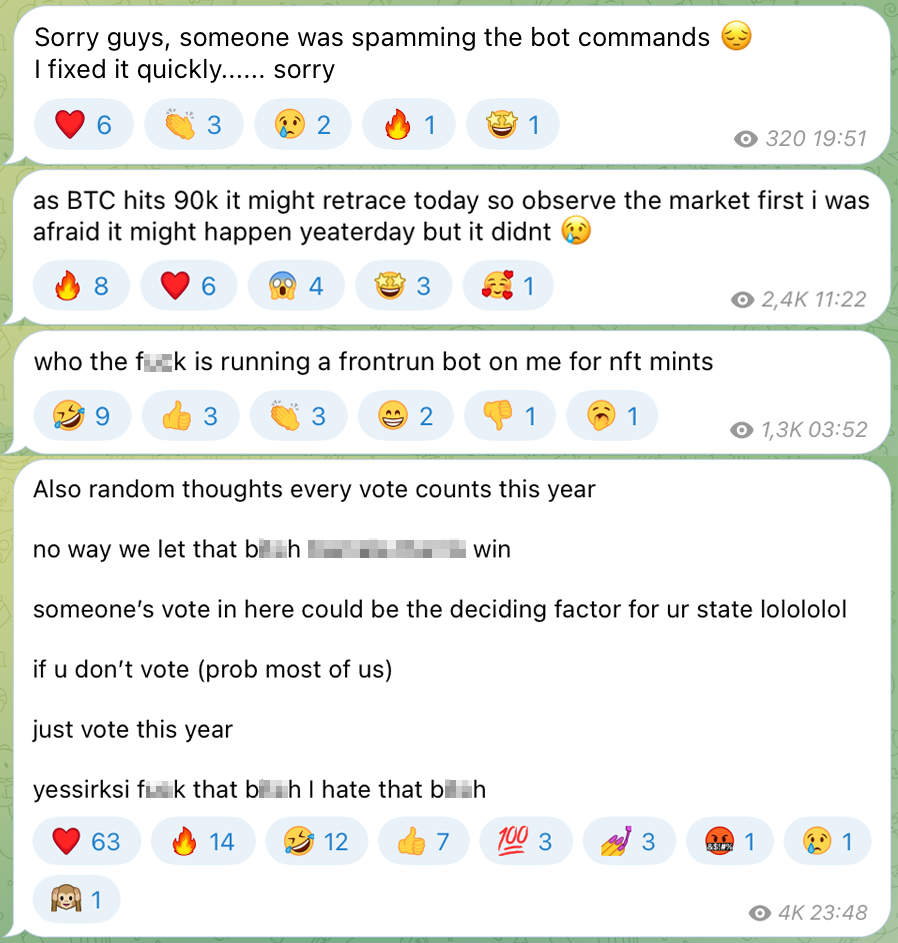}  
    \caption{Examples from our Telegram dataset showing that messages expressing negative or neutral emotions receive predominantly positive reactions.}
    \label{fig:tg-screenshot}
\end{figure}

To shed light on why joy behaves differently from other emotions, we examine the fine-grained emotional subtypes associated with each coarse emoji emotion category. Figure~\ref{fig:sankey} shows a Sankey diagram mapping each coarse emoji emotion to its fine-grained variants. Notably, while categories like anger or disgust map to narrowly defined emotional expressions, joy includes a wide range of sub-emotions, such as confidence, love, celebration, agreement, and gratitude. These variants often signal social approval, support, or encouragement, rather than shared affect. This broader functional range helps explain why joy reactions are frequently misaligned with the emotional content of messages, suggesting that they do not function as an emotional response but rather as a vehicle for social feedback. The mismatch becomes evident in real examples from the dataset, shown in Figure~\ref{fig:tg-screenshot}, where joy reactions are frequently used in response to emotionally negative messages such as those expressing sadness, fear, or anger. For example, reacting with a heart emoji to a sad post might mean ``I'm here for you'' more than ``I feel love,'' reflecting group dynamics and social norms. 

\subsection{RQ3 - Prediction of emoji reactions}

We employ a Random Forest classifier with class balancing, achieving a moderate yet consistent performance, with accuracy $= 0.71$ and macro-F1 $= 0.71$. This provides a solid base for further interpreting results through SHAP, which estimates the marginal impact of each input feature on the predictions of the model.

\begin{figure}[t]
  \includegraphics[width=\columnwidth]{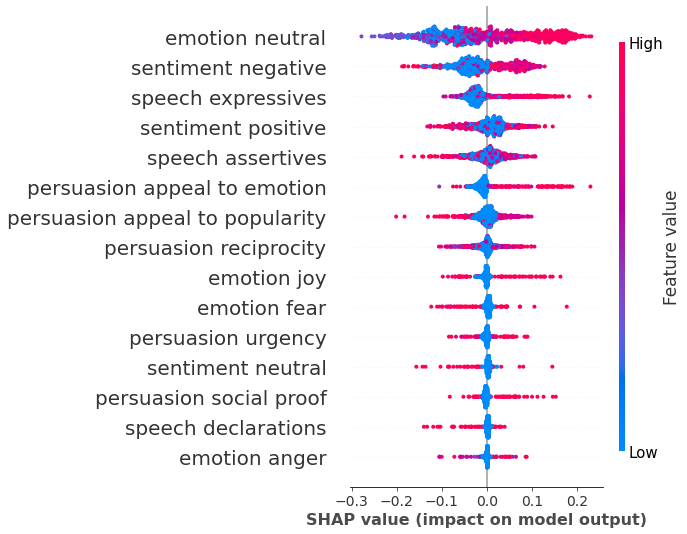}
  \caption{Impact of linguistic features on predicting high emoji reaction engagement. 
  }
  \label{fig:shap-values}
\end{figure}

Figure~\ref{fig:shap-values} shows the top $15$ contributing features (red-colored points represent high, while blue low) and their influence on the prediction toward high engagement. The top predictors of emoji reactions include Emotion Neutral, Sentiment Negative, and (to a lesser extent) Sentiment Positive, suggesting that the emotional valence of a message, whether positive or negative, plays a key role in shaping reaction dynamics. Interestingly, Emotion Neutral emerges as the strongest individual predictor, indicating that even emotionally flat messages can elicit strong social responses. 
Notably, stronger negative sentiment correlates with higher reaction rates, while less intensely positive sentiment is also linked to higher reactions. This pattern suggests that both negative emotional content and muted positivity may drive greater engagement, possibly because they elicit stronger social responses or invite interpretation.
These findings align with prior research on negativity bias in online platforms, which shows that negatively valenced messages tend to attract more interaction on social platforms~\cite{schone2023negative,watson2024negative}. Our novel results indicate that this dynamic extends to emoji reactions as well.

Linguistic categories also contribute. For example, Speech Expressives and Speech Assertives are among the most impactful predictors, suggesting that stylistic elements of the message may influence emotional engagement beyond what is said. Several persuasion strategies such as Appeal to Emotion, Appeal to Popularity, and Reciprocity also emerge as relevant predictors. In contrast, fine-grained emotion labels such as joy, anger, or fear showed comparatively modest influence.

\section{Discussion and Conclusions}

Our analysis reveals that emoji reactions on Telegram do not reliably act as emotional mirroring and may instead signal social support. While most prior work has focused on emojis embedded in text, showing that they amplify sentiment~\cite{shiha2017effects}, enhance sentiment classification~\cite{liu2021improving}, and provide valuable emotional context~\cite{lou2024emoji, khan2025sentiment}, this view has also been implicitly extended to emoji reactions, treating them as proxies for sentiment~\cite{pool2016distant}. Our findings challenge this assumption. Through ablation analysis, we show that joy is the most disruptive emotion in terms of alignment. Unlike other categories, joy reactions frequently appear in response to emotionally negative or neutral messages. This suggests that joy serves a pragmatic role, signaling approval, support, or social bonding. These patterns indicate that emoji reactions on Telegram may operate more as social cues than as emotional reflections. 
This urges caution against using emoji reactions as ground truth for sentiment analysis. This also raises concerns for AI models that might estimate public opinion based on emoji reactions.
Our results also help explain why negative or controversial content online often receives higher engagement, as reactions may reflect support for the person posting, complementing work by~\citet{morales2025thumbs}, where authors highlighted positive emoji reactions in the context of toxic discourse as approval signal. Moreover, we show that this bias toward positive reactions holds across a wider range of communicative intents and emotional tones. 
Finally, we show that stronger negative sentiment and muted positivity are linked to higher emoji reaction rates, extending the negativity bias observed in online platforms~\cite{schone2023negative,watson2024negative} to emoji-based interactions.
As our analysis focuses on crypto-related messages, a domain where emoji reactions are dense, public, and central to community interaction, the findings describe how emoji reactions function in this specific communicative context, while also pointing to the need to examine whether the same patterns appear in other Telegram communities or online platforms.
In conclusion, we provide insights into the social meaning of emoji reactions on Telegram, suggesting that they serve a distinct communicative purpose compared to in-text emojis. Recognizing this distinction is especially important for research on online engagement, misinformation, political discourse, where emoji reactions can be used as signals or indicators of emotional resonance or public sentiment.

\section{Limitations}

First, our findings rely on automated classifiers, including sentiment and emotion models based on BERT or gpt-4o, which have not been specifically validated on our dataset and may therefore introduce the usual degree of uncertainty associated with automated annotations.
Second, as our dataset consists mostly of public cryptocurrency-related Telegram channels, the communicative dynamics we observe may reflect domain-specific norms, which limits the generalizability of our findings and highlights the need to examine other contexts. Doing so would also clarify our interpretation of the positivity bias: although we view it as consistent with social-approval, engagement, or banter signaling, other explanations remain possible, including community norms specific to this domain that may encourage more positive reactions. Nonetheless, evidence that emoji reactions do not reliably reflect emotional resonance, at least in this setting, remains informative. At the same time, our analytical approach is general and can be applied to other domains and platforms.
Beyond these interpretative factors, the Telegram reaction system itself introduces structural constraints. The set of standard emoji reactions is finite, and our study does not consider custom emoji sets. In addition, not all channels enable reactions, and when they do, administrators may restrict the available options to a small and often predominantly positive set. These features can introduce selection bias and promote approval-oriented responses by design. Future work could extend these insights by examining how reaction emojis operate within a broader range of digital communication practices.

\bibliography{custom}

\end{document}